\documentclass[twocolumn,amsmath,amssymb,pre]{revtex4}
\usepackage{graphicx,color,longtable}

\begin{document}

\makeatletter

\title{Structure and kinetics in the freezing of nearly hard spheres}

\author{Jade Taffs}
\affiliation{School of Chemistry, University of Bristol, Bristol, BS8 1TS, UK.}

\author{Stephen R. Williams}
\affiliation{Research School of Chemistry, Australian National University, Canberra, ACT 0200, Australia.}

\author{Hajime Tanaka}
\affiliation{Institute of industrial science, The University of Tokyo, 4-6-1 Komaba, Meguro-ku, Tokyo, 153-8505, Japan.}

\author{C. Patrick Royall}
\affiliation{School of Chemistry, University of Bristol, Bristol, BS8 1TS, UK.}

\begin{abstract}
We consider homogeneous crystallisation rates in confocal microscopy
experiments on colloidal nearly hard spheres at the single particle
level. These we compare with  Brownian dynamics simuations by carefully modelling the softness in the interactions with a
Yukawa potential, which takes account of the electrostatic charges present
in the experimental system. Both structure and dynamics of the colloidal fluid are very well matched between experiment and simulation, so we have confidence that the
system simulated is close to that in the experiment. In the regimes
we can access, we find reasonable agreement in crystallisation rates
between experiment and simulations, noting that the larger system size
in experiments enables the formation of critical nuclei and hence crystallisation at lower supersaturations than the simulations.
We further examine the structure of the metastable
fluid with a novel structural analysis, the topological cluster classification.
We find that at densities where the hard sphere fluid becomes metastable,
the dominant structure is a cluster of $m=10$ particles with five-fold
symmetry. At a particle level, we find three regimes for the crystallisation process: 
metastable fluid (dominated by $m=10$ clusters), crystal and a transition region
of frequent hopping between crystal-like environments and other ($m\neq10$) structures.\end{abstract}

\maketitle

\section{Introduction}

\begin{figure}
\includegraphics[width=85mm]{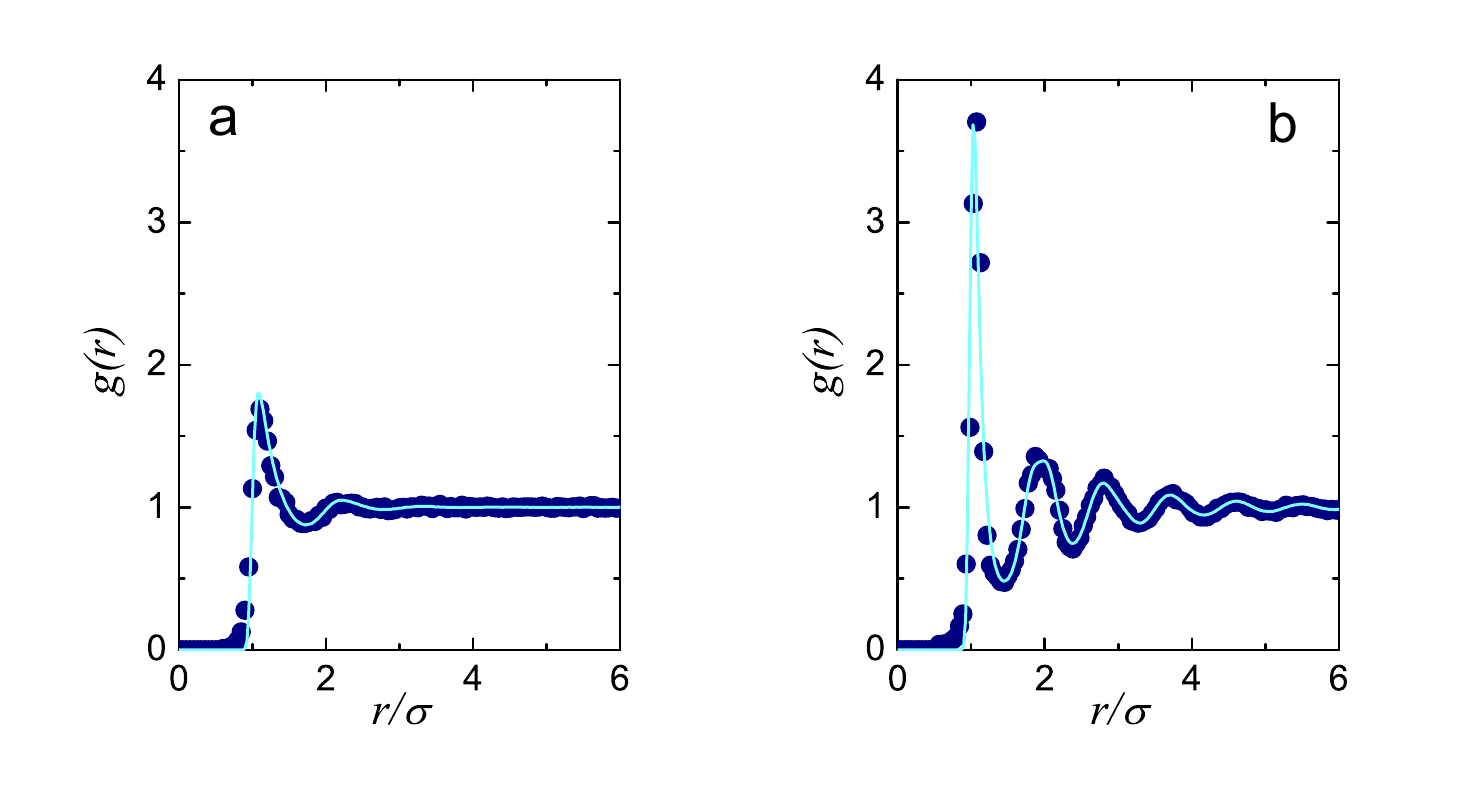} 
\caption{(color online) Radial distribution functions in experiment
and simulation. (a) low density, $\phi=0.27$, (b) higher density
$\phi=0.53$. In both cases, simulation parameters were carefully
adjusted to experiment. Line, simulation data, circles, experimental
data.}
\label{figG} 
\end{figure}

Crystallisation is a long-standing challenge, due not least to
its local nature, where rare events on microscopic time- and length-scales
initiate the phase transition ~\cite{sear2007}. This lack of understanding
of crystallisation can have very significant practical consequences,
for example in control of drug production ~\cite{morissette2003}.
It appears challenging to make much progress with conventional materials,
due to the local nature of nucleation events which lead to crystallisation, however particle-resolved studies of model systems such as colloidal dispersions which capture
the essential thermodynamics provide the necessary detail required \cite{book}.

Colloidal `hard' spheres are important in the understanding
of crystallisation. Few systems have received so much attention, not least
because both simulations and experiments can access relevant timescales
and particle-level structural lengthscales ~\cite{auer2001,auer2001a,gasser2001,aastuen1986,harland1997}. The general phenomenology
of hard sphere crystallisation has been well established for a decade
\cite{palberg1999}: at low supersaturations, close to the hard sphere
freezing transition at a volume fraction of $\phi_{f}=0.494$ ~\cite{hoover1968}, 
crystallisation is dominated by rare events leading to the
formation of large nuclei. Higher supersaturation leads to a very
strong rise in nucleation rate, and upon increasing the volume fraction,
approaching the hard sphere glass transition, crystallisation has
been observed at times less than the structural relaxation time ~\cite{zaccarelli2009xtal,sanz2011},
while at higher volume fractions still ($\phi=0.62$), crystallisation
is not seen on the experimental timescale. Despite this phenomenlogy,
very large discrepancies have been found in nucleation rates predicted
by simulation using biased ensemble averaging and experiment ~\cite{auer2001,auer2001a,auer2004},
which remain unexplained ~\cite{filion2011,schilling2011}.
Neither the inclusion of polydispersity ~\cite{auer2001a,auer2004} nor electrostatic charge
~\cite{auer2002} in the simulations has resolved this situation, although the former
has been shown to have profound and complex consequences for nucleation ~\cite{schope2006,schope2007}.

Charles Frank originally suggested that 13-membered icosahedra might suppress crystallisation in the Lennard-Jones system ~\cite{frank1952}, and recently there has been a resurgence of interest in the role of local structure in crystallisation. In simulations of hard spheres, five-fold symmetry has been identifiedboth with the suppression of
crystallisation ~\cite{karayiannis2011,karayiannis2012} and found at the
centre of crystal nuclei ~\cite{omalley2003}. Locally dense amorphous crystal
precursors have been identified in the metastable hard sphere fluid ~\cite{schilling2010} and have also been found in softened systems ~\cite{lechner2011,russo2012}.
One of us identified a mechanism for crystallisation through increased crystal-like
ordering in the fluid prior to the formation of a nucleus, thereby
lowering the free energy barrier \cite{kawasaki2010pnas,kawasaki2011corr,kawasaki2010jpcm}, and that this entails no change in local density ~\cite{russo2011}.
It was also shown that in weakly size-asymmetric binary hard
sphere systems, crystallites can form quickly, but apparently become
`poisoned' ~\cite{williams2008}. 

Pioneering particle-resolved experiments ~\cite{gasser2001} identified local structure, and more recent experiments on `hard' spheres too polydisperse to crystallise have shown a degree of fivefold symmetry which, along with local crystalline order, has been related to slow dynamics ~\cite{leocmach2012}. Here we consider local structure in crystallisation in a particle-resolved colloidal model system.
While such experimental studies can in principle resolve
mechanisms of crystallisation, quantitative comparison
to simulation and theory is very challenging, due to the limited accuracy
with which colloidal volume fractions can be measured ~\cite{poon2012},
combined with the lack of control over (and often knowledge of) interparticle interactions
upon which crystallisation rates critically depend ~\cite{auer2002,auer2003,royall2012myth}.
Quantitative agreement between experiment and simulation has been
obtained in the case of \emph{heterogeneous} crystallisation of nearly hard spheres, initiated
by a wall, where the crystallisation rate is less sensitive to the
volume fraction compared to homogenous crystallisation ~\cite{sandomirski2011}.

%It was also shown that crystal-like bond orientational order in a liquid not only 
%helps crystal nucleation \cite{kawasaki2010pnas}, but also selects crystal polymorph to be %nucleated \cite{russo2012}. 
%Roles of icosahedral order in slow relaxation and crystallization have also been 
%discussed on the basis of both experiments \cite{leocmach2012} and simulations %~\cite{russo2011}. 

\begin{figure}
\includegraphics[width=45mm]{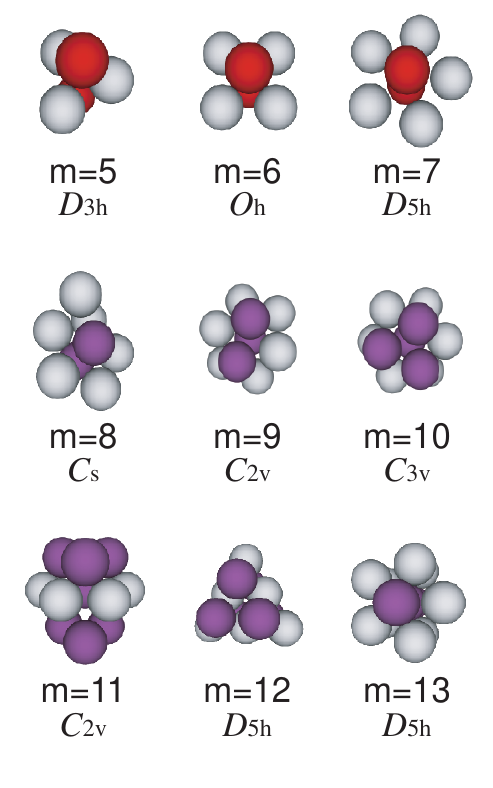}
\caption{(color online) Clusters detected by the topological cluster classification.
These structures are minimum energy clusters of the Morse potential with $\rho_0=25.0$
\cite{doye1995}.}
\label{figTCC} 
\end{figure}

Here we present a careful comparison of experiment and simulation
in a system of nearly hard spheres which undergo homogenous nucleation. We interpret our results with a novel structural method, the topological cluster classification
(TCC) ~\cite{williams2007,royall2008,malins2012}, which directly identifies a number of local
structures. Our mapping between experiment and simulation reveals 
good agreement in crystallisation rates at the range of supersaturation we accessed. 
%At lower supersaturation, experiments crystallise more readily than simulations. 
We find that the metastable fluid is dominated by 10-membered fivefold symmetric structures reminiscent
of the 13-membered icosahedra proposed long ago as a mechanism for
the suppression of crystallisation ~\cite{frank1952}. 

\begin{figure}
\includegraphics[width=90mm]{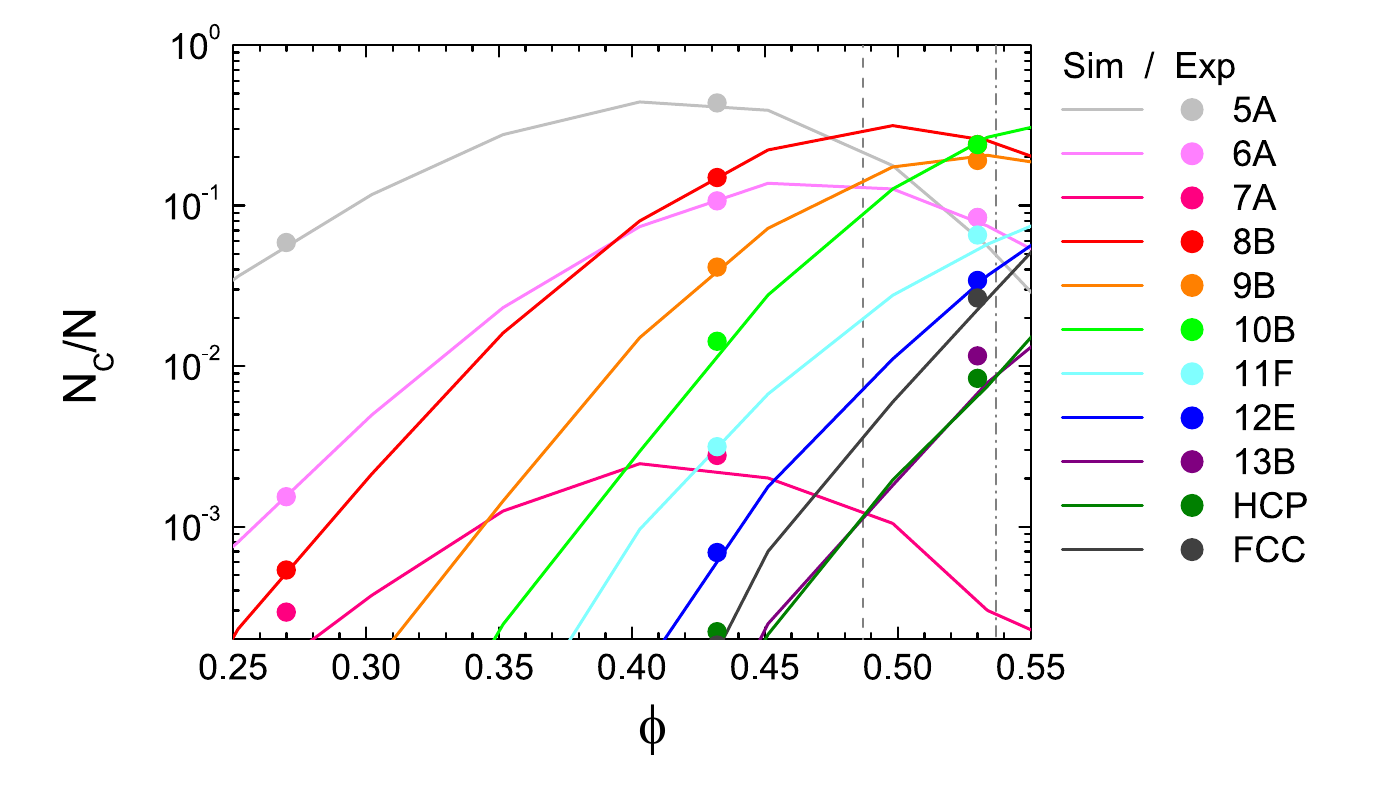} 
\caption{(color online) Structural changes upon increasing density in the
nearly-hard sphere fluid. Lines are simulation, according to eq. \ref{eqTMR},
circles are experiment. Data for metastable fluids (some of which subsequently
crystallise) are taken at times $\ll\tau_{x}$ . Dashed lines are estimated freezing and 
melting volume fractions for our system, as described in section \ref{subsectionMapping}.
\cite{hynninen2003}.  }
\label{figTCCrho} 
\end{figure}

\section{Methods}

\subsection{Experimental}

We used polymethyl methacrylate (PMMA) particles of diameter $2.00$
$\mu$m with a polydispersity of $4.0$ \% as determined by static
light scattering. This degree of polydispersity is insufficient to
have much impact on phase behaviour \cite{sollich2010}. Swelling of the colloids cannot
be ruled out in the density and refractive index matching solvent
mixture of cis-decalin and cyclohexyl bromide used \cite{poon2012}. However,
as far as crystallisation is concerned, 
electrostatic charge, which is not entirely screened by the tetrabutyl
ammonium salt added contributes a further degree of uncertainty in determining the effective
colloid volume fraction. If ignored, the effects of electrostatics are
quite sufficient to leave measures of crystallisation rates quantitatively meaningless
\cite{royall2012myth,auer2002}. For this reason we map simulations
carefully to the experiments. We use confocal microscopy (Leica SP5) to track
the particle coordinates. Heterogenous nucleation is prevented by
weakly sintering larger (3.5 $\mu$m) polydisperse colloids onto
the wall of the sample cell. We imaged at least 50 $\mu$m from the
wall and saw no sign of heterogeneous crystallisation.

\subsection{Mapping simulation to experiment}
\label{subsectionMapping}

Crystallisation experiments are compared with standard Brownian dynamics (BD) simulations, with a system size of N = 2048 and 10976 particles and a timestep of 0.1 simulation time units. Fitting of the  experimental radial distribution function is carried out using Monte Carlo (MC) simulations with N = 2048 particles. Both the BD and MC simulations are carried out in the canonical (NVT) ensemble. Our particle-resolved experiments enable an innovation: simulations take as their starting point experimental coordinates sampled from a fluid at a time small compared to that required for crystallisation. These are treated with $160$ MC sweeps to remove small overlaps resulting from coordinate tracking errors. Particle interactions are modelled with a truncated Morse potential with a Yukawa component, which approximate the hard core and electrostatic charging  of the colloidal particles respectively.  

\begin{eqnarray}
\beta u(r)= \beta \varepsilon_{TM} \left[1+e^{\rho_{0}   \left(1-r/\sigma_{ij} \right)}  \left(  e^{\rho_{0} \left( 1- r/\sigma_{ij}   \right)} -2  \right) \right] 
\nonumber \\ 
+ \beta\varepsilon_{Y}\frac{e^{-\kappa(r/\sigma_{ij}-1)} }{r/\sigma_{ij}}  
\label{eqTMR}
\end{eqnarray}

\noindent Here the shifted Morse component (left term) is truncated at $r=\sigma_{ij}$ (where it vanishes)
and the Yukawa component (right term) is truncated at $r=2 \sigma_{ij}$.
$\beta=1/k_{B}T$ the thermal energy, $\sigma_{ij}$
is the mean of the diameters of colloids $i$ and $j$ and $r$ is the
center separation. The truncated Morse potential is fixed with strength $\varepsilon_{M}=1.0$ and range parameter $\rho_{0}=25.0$. The contact potential of the Yukawa
contribution $\beta\varepsilon_{Y}=Z^{2}\lambda_{B}/[(1+\kappa\sigma/2)^{2}\sigma]$.
Here $Z$ is the number of charges on the colloid and $\lambda_{B}$ is the Bjerrum length. The inverse Debye screening length is denoted by $\kappa=\sqrt{4\pi\lambda_{B}\rho_{ion}}$
where $\rho_{ion}$ is the number density of (monovalent) ions.

We fix the Yukawa parameters to the expermental data. Our approach
follows \cite{royall2003,royall2006}, where the Yukawa interaction
parameters $\beta\varepsilon_{Y}$ and $\kappa\sigma$ are adjusted
such that the experimental radial distribution function is well reproduced
by the simulation (Fig. \ref{figG}). In this case we find $\kappa\sigma=30.0\pm5.0$
and $\beta\varepsilon_{Y}=1.0\pm0.25$, which corresponds to a Debye length of 67 nm (or an ionic strength of 1.4 $\mu$M) and colloid charge of $Z=200$. These are comparable
to previous work on similar systems \cite{yetiraj2003,royall2003,royall2006,royall2007,campbell2005}.

We treat polydispersity with a Gaussian distribution in
$\sigma$ with 4\% standard deviation (the same value as the size
polydispersity in the experimental system). The radial distribution function of each experimental
state point was fitted for a (metastable) fluid with MC simulation. We then quote the state point in units of $\phi=V_{part}/V_{box}$ where the volume of the particles is taken as 
$V_{part}=\frac{\pi }{6}\sum _{i}^{N}\sigma _{i}^{3}$.
%$\phi=6 \rho / \pi$ where $\rho$ is the number density used in the simulation. Unless otherwise stated, all simulation results correspond to a system with 4\%  polydispersity.

There are a variety of ways
to estimate the freezing transition for a system of weakly repulsive
spheres. Among the more accurate ~\cite{royall2012myth} appears to
be to interpolate exact simulation results for hard-core Yukawa systems
~\cite{hynninen2003} with the hard sphere values. This yields volume fractions for freezing and melting for a 
weakly charged system. Since we use a slightly softened core here, we estimate the impact of this softening on the phase behaviour by calculating the Barker-Henderson effective hard sphere diameter $\sigma_{eff}$
\begin{equation}
\sigma_{eff}=\intop_0^\infty dr\left[1-\exp\left(-\beta u(r)\right)\right],
\label{eqBH}
\end{equation}
\noindent where $u(r)$ is the interaction potential, i.e. Eq. (\ref{eqTMR}) or a hard core with the same Yukawa term. The effective hard sphere diameters are $\sigma_{eff}^{HCYUK}=1.021\sigma$ and $\sigma_{eff}^{TMYUK}=1.018\sigma$ for the hard-core Yukawa and the truncated Morse system we use here. To approximately
include the slight effect of the core softening, we scale the volume fractions for freezing and melting by $(\sigma_{eff}^{HCYUK}/\sigma_{eff}^{TMYUK})^3$. The
core softening then leads to a change of around $0.004$ in $\phi$ in addition to the effect of
the Yukawa repulsion. Thus for our system, we estimate the freezing volume fraction $\phi_{f}=0.487$ and melting $\phi_{m}=0.537$.
Note that we neglect the effect of the 4\% polydispersity, whose impact on the freezing and melting  volume fractions we expect to be slight \cite{sollich2010}.  

\begin{figure}[H]
\includegraphics[width=80mm]{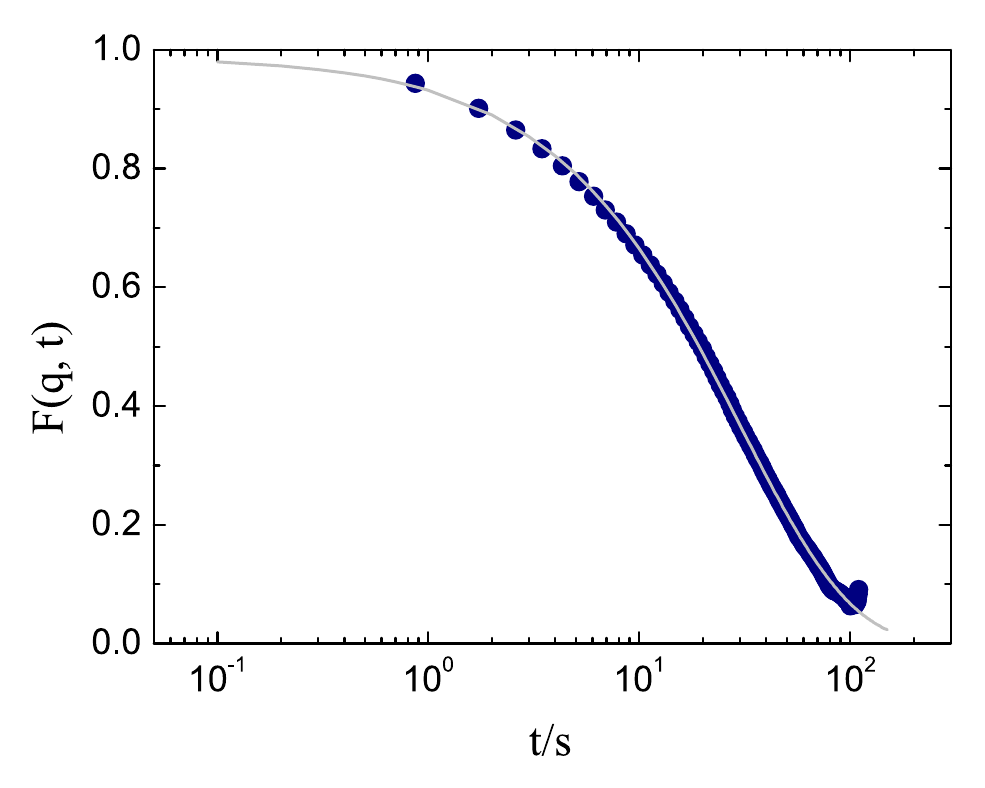} 
\caption{Intermediate scattering function for experimental data at $\phi=0.43$.
The wavevector is taken close to the first peak in the static structure factor. Grey line is a
streched exponential fit (see text for details).
\label{figISF}} 
\end{figure}

\begin{figure}[H]
\includegraphics[width=80mm]{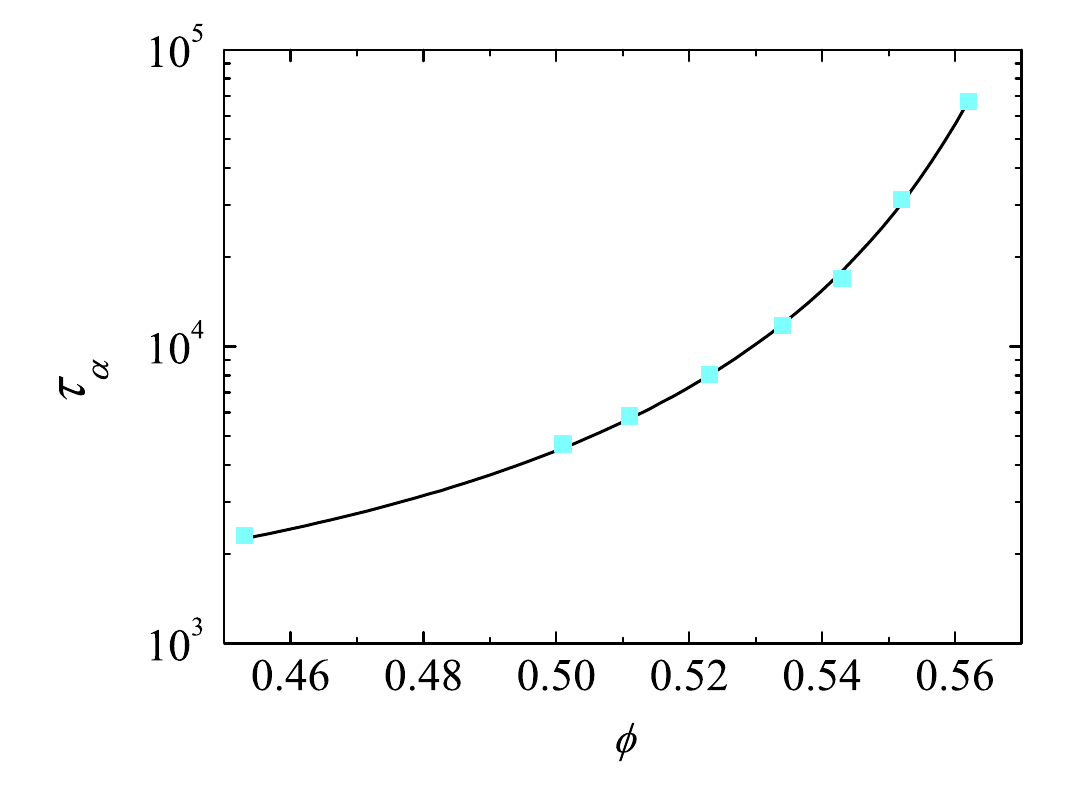} 
\caption{(color online) Structural relaxation time as a function of $\phi$. Solid line is a Vogel-Fulcher-Tammann fit (see text for details).}
\label{figAngell} 
\end{figure}

\subsection{Structural Analysis - Topological Cluster Classification}
\label{subSectionStructural}

To analyse the structure, we identify the bond network using a maximum bond length of $1.4 \sigma$ and a modified Voronoi construction ~\cite{williams2007}. For bond lengths
greater than $1.4\sigma$, the network in condensed systems is insensitive to the bond length. Having identified the bond network, we use the Topological
Cluster Classification (TCC) to determine the nature of the 
local environment of each particle~\cite{williams2007,malins2012}.
This analysis identifies all the shortest path three, four and five
membered rings in the bond network. We use the TCC to find structures topologically identical to clusters which are global energy minima of the Morse potential for the range we consider $\rho_{0}=25.0$, as listed in
~\cite{doye1995} and illustrated in Fig. \ref{figTCC}. 

Now the system we consider interacts not via a full Morse potential, rather our truncation takes the repulsive component only, in a similar spirit to the approach Weeks, Chandler and Andersen (WCA) used for the Lennard-Jones model ~\cite{weeks1971}. While that approach is well-known to reproduce accurately the fluid structure at the pair level, one might expect deviations for higher-order structure such as that probed by the TCC. In fact we found that for short-ranged systems, clustering is \emph{enhanced} in the case of truncation ~\cite{taffs2010jcp}. Unlike many analyses, for example those which use bond-orientational order parameters ~\cite{steinhardt1983}, our emphasis on bond topology distinguishes between icosahedra and the 13-membered $D_{5h}$ structure illustrated in Fig. \ref{figTCC} which is the minimum energy cluster for the Morse potential with $(\rho_{0}=25.0)$. We have also checked for
the icosahedron and found small quantities ($\lesssim 1\%$)~\cite{taffs2010jcp}. In addition we identify the thirteen particle structures which correspond to 
FCC and HCP in terms of a central particle and its twelve nearest neighbours. 
For more details see ~\cite{williams2007,malins2012}. 
If a particle is a member of more than one cluster, we take it to reside in the larger cluster.

\begin{figure*}[H]
\includegraphics[width=160mm]{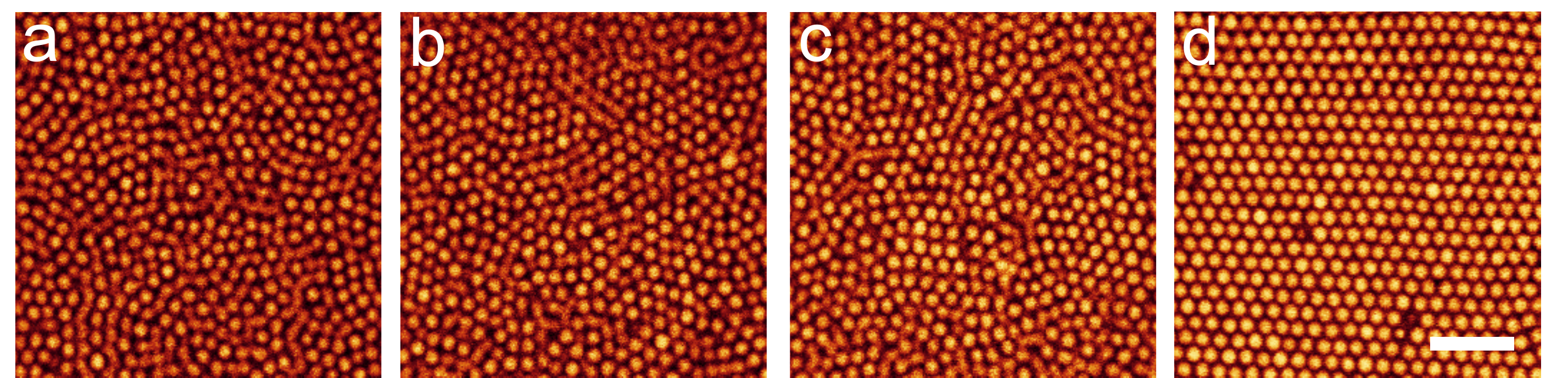} 
\caption{(color online) Confocal microscopy images of crystallisation in nearly-hard spheres for $\phi=0.54$.
(a) 600 s  (2.3 $\tau\alpha$), (b) 4500 s ( 17.4 $\tau\alpha$), (c) 7200 s ( 27.9 $\tau\alpha$)
(d) 81900 s ( 316.9 $\tau\alpha$)
bar=10 $\mu$m.} 
\label{figPix} 
\end{figure*}

\begin{figure*}[H]
\includegraphics[width=160mm]{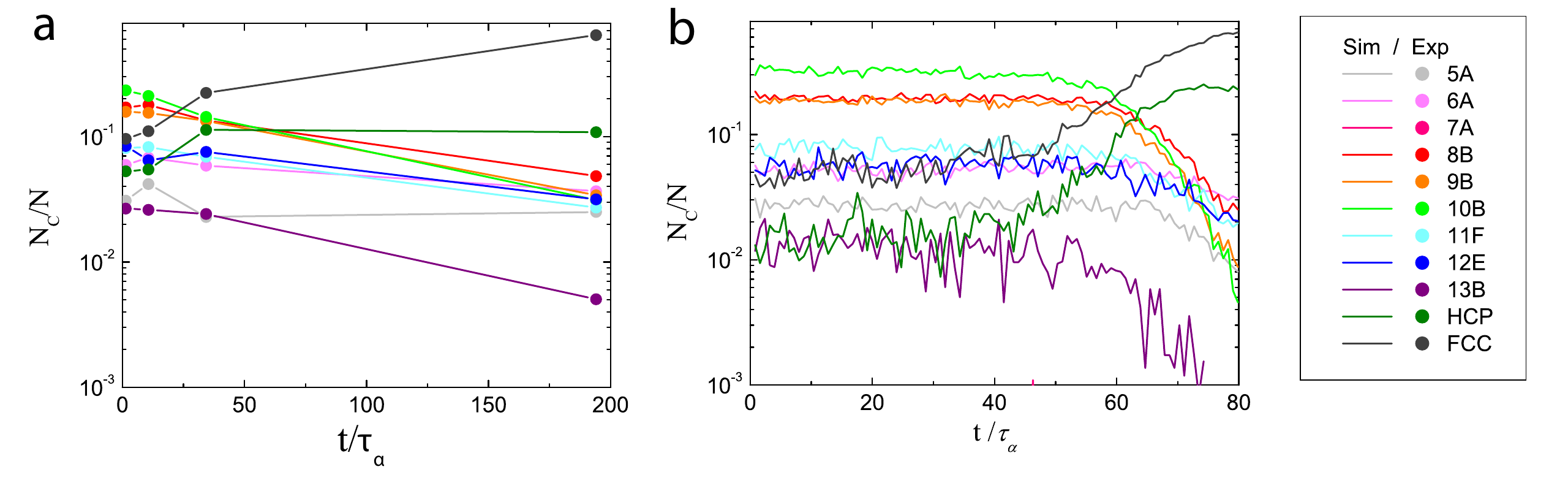} 
\caption{(color online) Topological cluster classification analysis of crystallisation,
experimental (a) and simulation data (b). Here $\phi=0.56$.
\label{figTCCtime}} 
\end{figure*}

\section{Results}

\subsection{Fluid structure: matching state point}

We begin our presentation of results by considering the stable and
metastable fluids. For metastable fluids, the cluster populations
are sampled from the period prior to crystallisation (see next section).
The topological cluster classification (TCC) analysis shows a considerable increase
in cluster populations as a function of volume fraction, as shown
in Fig. \ref{figTCCrho}. Some smaller clusters are subsumed into larger
clusters for $\phi\gtrsim0.55$. We see a sharp rise in the fivefold
symmetric 10-membered $C_{2v}$ cluster we term `10B' following 
\cite{doye1995}, such that by $\phi\gtrsim0.54$, it is
the dominant cluster in the fluid. We estimate freezing in our nearly hard sphere system at $\phi=0.487$ as noted in section \ref{subSectionStructural}.
We note that the experimental
and simulation data are very well-matched in Fig. \ref{figTCCrho}. This
gives us confidence that the fluid structure of the experiments is
accurately reproduced in simulation, in other words that the state
point is well matched and that we are therefore in a strong position
to investigate any possible discrepancies between the experimental
and simulated system.

\begin{figure*}[H]
\includegraphics[width=160mm]{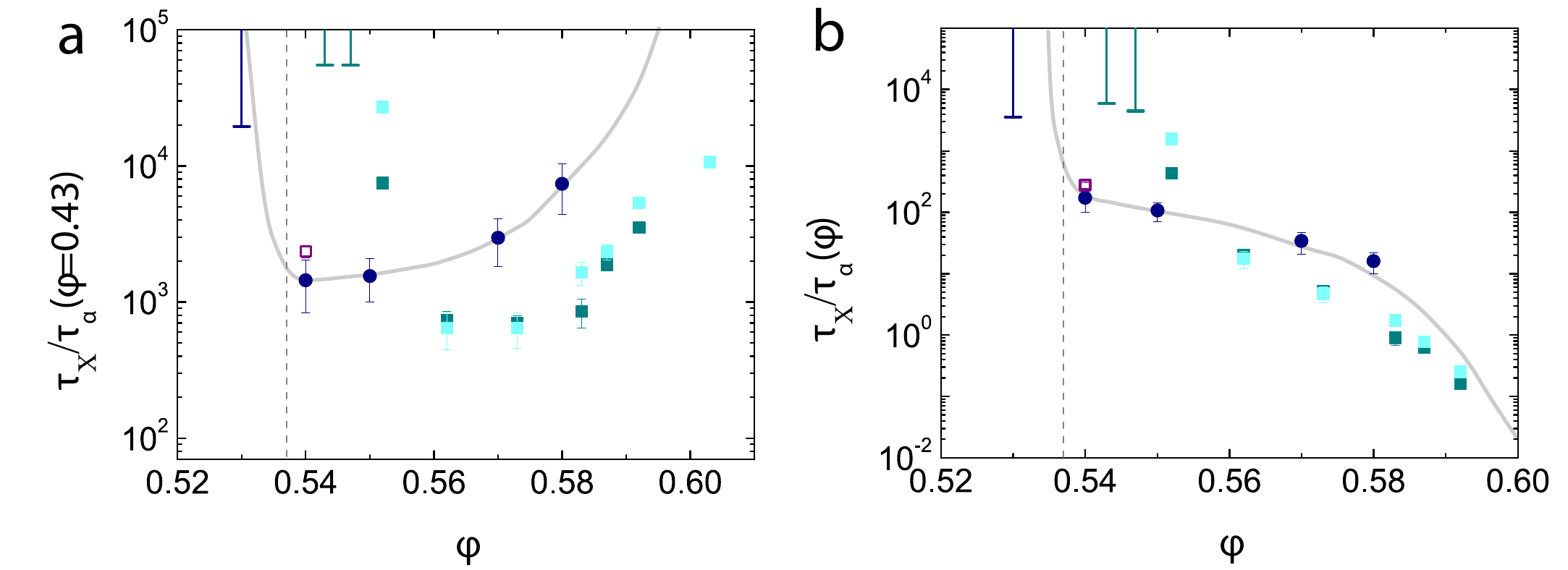}
\caption{(color online) Crystallisation times in terms of $\tau_\alpha(\phi=0.43)$ (a) and $\tau_\alpha(\phi)$. Circles are experimental data, dark and light squares are simulation data for polydisperse systems of $N=10976$ and $N=2048$ respectively. Unfilled square is for a monodisperse system with $N=10976$.
Dashed lines are melting estimated as described in section \ref{subsectionMapping}. Solid lines are to guide the eye. Error bars extending upwards are lower bounds for crystalisation times determined from experiments (light lines) and simulations (dark lines) which did not crystallise.}
\label{figXtalTime} 
\end{figure*}

\subsection{Matching timescales}

The timescales in the experiments and simulations are matched as follows. 
The dynamics slow upon increasing $\phi$ and we define a structural relaxation time from the intermediate scattering function (ISF). We determine
the intermediate scattering function (ISF) $F(\mathbf{k},t)=\langle \cos(\mathbf{k} \cdot (\mathbf{r}(t+t')-\mathbf{r}(t'))^{2}) \rangle$
where the wavevector \textbf{$\mathbf{k}$} is taken close to the
main peak in the static structure factor and $\mathbf{r}$ is the
location of each particle at time $t$ and the angle brackets denote
a statistical average. An ISF is shown in Fig. \ref{figISF}. 
We fit the tail of the ISF with a stretched exponential $F(\mathbf{k},t)=C\exp[(-t/\tau_{\alpha})^{\beta}]$
to obtain the structural relaxation time $\tau_{\alpha}$.
We obtain $C = 0.99,$ $\tau_{\alpha} = 30.5$ s, $\beta = 0.8134$. 
$\tau_{\alpha}$ is then plotted as a function of $\phi$ in Fig. \ref{figAngell}. The data are well described
by a Vogel-Fulcher-Tammann (VFT) relation {$\tau_{\alpha}=\tau_{0}\exp[D\phi/(\phi_{0}-\phi)]$
where $\tau_{0}$ is a relaxation time in the normal liquid, $D$
is the `fragility index', and $\phi_{0}\approx0.616$ is the ideal glass transition volume fraction
 \cite{kawasaki2010jpcm}. We assume the scaling of the dynamics at this volume fraction
is the same for experiment and simulation. We equate the experimental and simulation structural relaxation 
times at $\phi=0.43$. Thus both experiment and simulation $\tau_\alpha$ are taken from the VFT fit (Fig. \ref{figAngell}).

\subsection{Structure in crystallisation}

The process of crystallisation is shown in Fig. \ref{figPix}. Here
$\phi=0.54$. Note that some small crystalline regions are present at $t=0$. We have previously demonstrated that even stable fluids have populations of particles in crystalline environments \cite{taffs2010jcp},
as shown in Fig. \ref{figTCCrho}, this rises markedly around the freezing
transition. These images suggest that the metastable fluid shows relatively
little change for $27.9\tau_{\alpha}\leq t${, but crystallisation subsequently
occurs at} $27.9\tau_{\alpha}\leq\tau_{x}\leq 316.9\tau_{\alpha}$. Henceforth
we define $\tau_{x}$ as the time at which $40\%$ of the particles
are identified in crystalline environments. Moderate changes to this threshold have no impact upon our conclusions.

\begin{figure*}[H]
 \includegraphics[width=160mm]{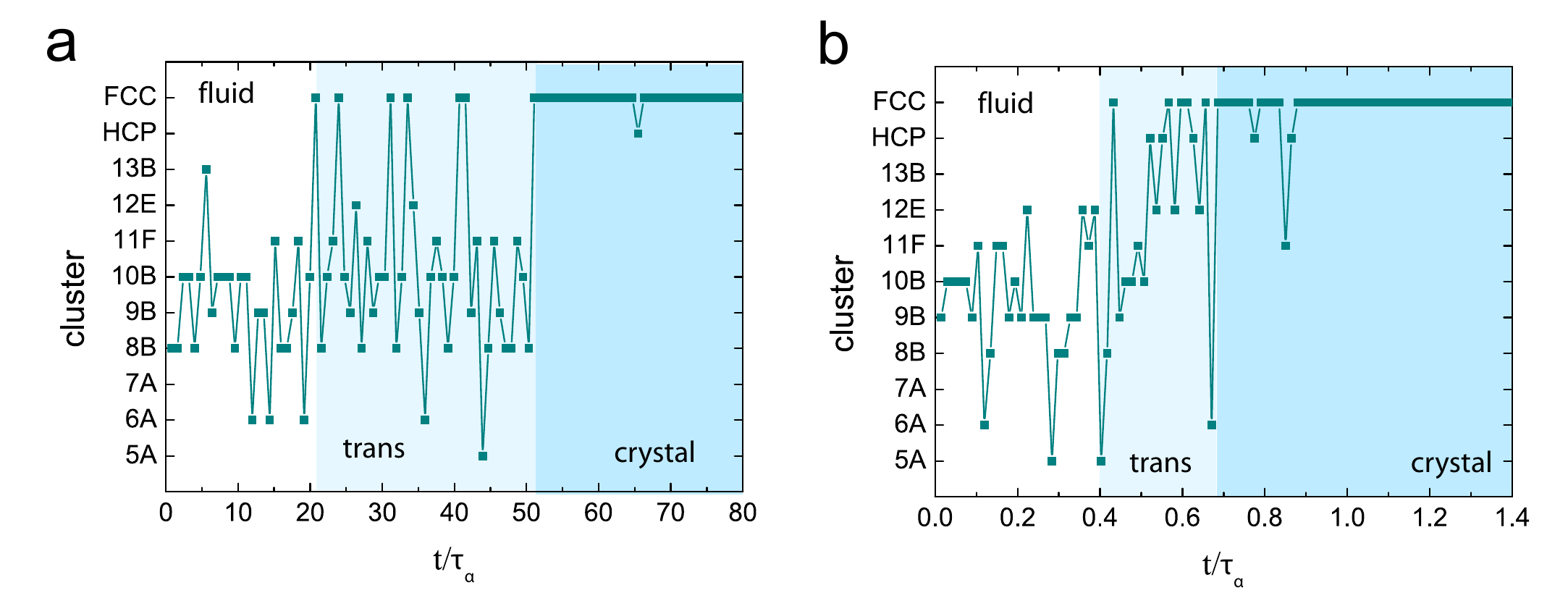} 
\caption{(color online) History of a single particle. (a) $\phi=0.55$,  (b) $\phi=0.58$.
Shaded areas mark the different regimes of fluid, transition and crystal, as described in the text. Data are shown from Brownian dynamics simulations.}
\label{figTimeSingle} 
\end{figure*}

\begin{figure*}[H]
\includegraphics[width=160mm]{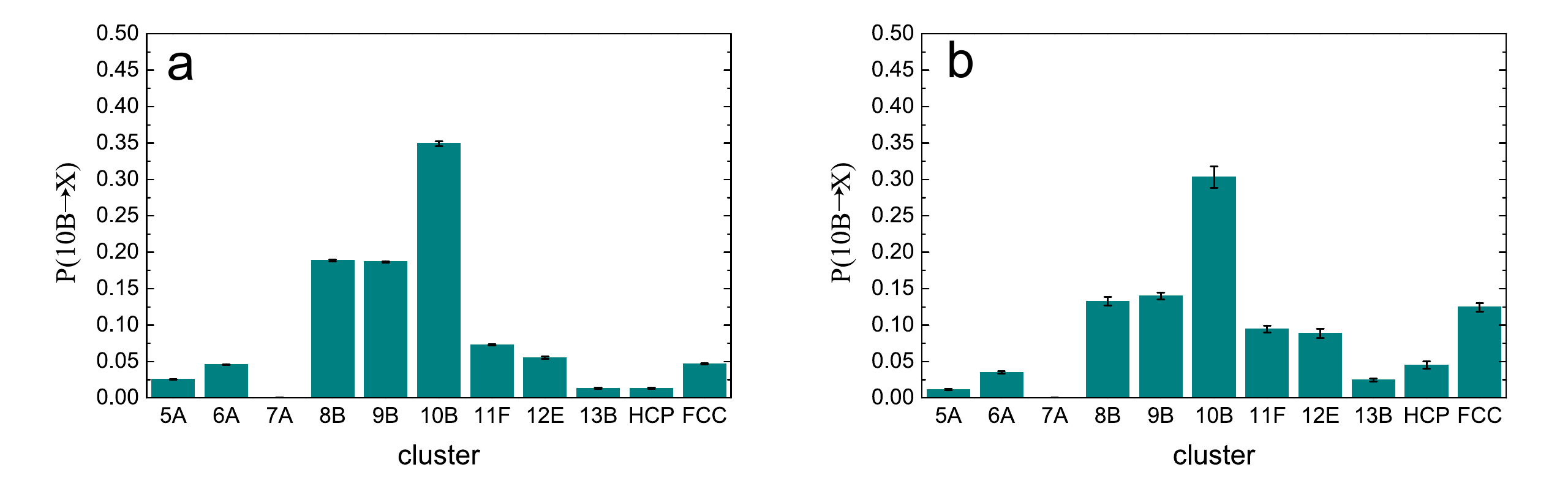} 
\caption{(color online) Transitions from the 10B cluster. (a) $\phi=0.55$, (b) $\phi=0.57$.
These are the probability for a particle to be found in a cluster at time $t+\tau_t$, given that it was in a 10B cluster at $t$. Here $\tau_t=0.08 \tau_\alpha$ for both (a) and (b). Data are shown from Brownian dynamics simulations.}
\label{figTransitions} 
\end{figure*}

We now turn to a TCC analysis of the crystallisation process. Figure
\ref{figTCCtime} shows the population in each TCC cluster as a function
of time for experiment (a) and simulation (b). In both experiment
and simulation, we identify three regimes. For $t\lesssim10\tau_{\alpha}$ in the experiment and 
$t\lesssim40\tau_{\alpha}$ in the simulation there is little change in cluster populations. 
%For $10\tau_{\alpha}\lesssim t\lesssim\tau_{x}$,
At intermediate times approaching $\tau_x$
(here $\tau_{x}=107$ and $64\tau_{\alpha}$ for this experiment and simulation run respectively), we see a steady growth in
particles identified in crystalline environments (predominantly FCC)
at the expense of particles in fluid environments. 
Most notable of the non-crystalline clusters is 10B, which drops
continuously throughout this period. At the level of this description,
then, crystallisation is interpreted as the conversion of 10B clusters
into FCC environments. At times larger than $\tau_{x}$, there is
a further decrease in non-crystalline clusters. However, note that,
on the timescale of these experiments and simulations, a reasonable
population (a few percent) of non-crystalline clusters remain at all
times.

\subsection{Comparision between experiment and simulation}

We now compare crystallisation times in simulation and experiment.
Recall that nucleation of `hard' spheres is found to exhibit strong
deviations between experiment and simulation ~\cite{auer2001,filion2011}.
We compare crystallisation times as shown in Fig. \ref{figXtalTime}. We see a reasonable
agreement for moderate values of $\phi\gtrsim 0.56$, but at lower supersaturation  $\phi \lesssim 0.55$ or $\phi-\phi_M \lesssim 0.01$, 
we find an emergent discrepancy between experiment and simulation. 
While no mapping between experiment and simulation is perfect ~\cite{poon2012,royall2012myth},
our careful analysis of state point and timescale leads us to believe that this discrepancy is not 
accounted for by a shift of $\phi$.
  
Now neither the simulations we employ here, \emph{nor the confocal
microscopy experiments} access the regime of low supersaturation
where the formation of a large nucleus is a rare event. In simulation, biasing techniques can
be employed, and while similar methodologies are in principle possible
in experiment \cite{hermes2011} the kind of precision required to
determine nucleation rates quantitatively remains some way off. An
important point then, is that unbiased simulation and confocal
microscopy experiment access similar regimes of supersaturation.
However the experiment has a rather larger system size than does the simulation.
The simulation box size is typically 2000 and 10,000 $\sigma^{3}$ for the $N=2048$ and 10976 system sizes respectively, while the experiments
are confined in capillaries of size $250\sigma\times250\sigma\times2500\sigma\sim 1.6\times10^{8}\sigma^{3}$. The imaging volume ($50\sigma\times50\sigma\times25\sigma\sim 6.3\times10^{5}\sigma^{3}$) is rather smaller than the whole system and crystals can nucleate outside this region (Fig. \ref{figPix}).
The rate of crystal growth has recently been determined in a very similar system \cite{sandomirski2011}, and the associated timescales are of order 100-1000$\tau_{\alpha}$, suggesting that for long times, crystals can spread throughout the sample, so the relevant volume is that of the entire system, rather than just the imaging volume. Thus, decreasing the supersaturation to $\phi-\phi_M \approx 0.005$, we see that experiments continue to crystallise, but for simulations, the time for crystallisation moves outside the accessible timescale. Note that, at higher supersaturation, nucleation rates increase strongly, so the crystallisation time $\tau_{x}$ is somewhat independent of system size.

While the argument presented above is physically attractive, we seek a more quantitative validation, and for this we turn to the results of 
Filion \emph{et al.} ~\cite{filion2011}. Clearly, for the system to crystallise, at least one nucleation event must occur. Taking the nucleation rate for the highest volume fraction in ref. ~\cite{filion2011}, for a monodisperse system $J=1.4 \times 10^{-5} \sigma^{3} \tau_B$ where $\tau_B$ is the time to diffuse a diameter in the dilute limit and is equal to $0.63 \tau_\alpha(\phi=0.43)$. The corresponding volume fraction relative to melting is $\phi-\phi_M=0.0051$. For our system, this corresponds to one nucleation event every $4.0$ and $21.6$ $\tau_\alpha(\phi=0.43)$ for $N=10976$ and $N=2048$ respectively. In other words, in our simulation timescales, we expect crystallisation.

However, the rate of nucleation is highly sensitive to \emph{polydispersity} ~\cite{auer2001a,auer2004}, which is $4\%$ here. Recall this is expected to have little effect on the equilibrium phase diagram \cite{sollich2010}.
While to the best of our knowledge, no precise predictions for nucleation rates in polydisperse systems have been made for the regime we access ($\phi \gtrsim \phi_M$), it seems reasonable to expect a shift of the nucleation rate as a function of $\phi$ by $\sim 0.015$ which has been found at lower supersaturation ~\cite{auer2001a,auer2004}. We note that the shape of the particle size distribution, rather than just its second moment, can be important ~\cite{schope2006,schope2007}, however here we shall assume that the effect of polydispersity is to effect a shift in $\phi$ of $0.015$ in the nucleation rate. In other words, we take the rate for $\phi-\phi_M=-0.01$ for a monodisperse system to apply for $\phi-\phi_M=0.005$ for our system. The nucleation rate at $\phi-\phi_M=-0.01$ is 
four orders of magnitude lower than that at $\phi-\phi_M=0.005$  ~\cite{filion2011}. At such low rates, we expect no nucleation in simulation, but the larger system size in experiment (conincidentally four orders of magnitude larger than the simulation) is sufficient for nucleation to occur on our timescales, as is consistent with the crystalisation that we see.

If this analysis is correct, for a monodisperse system at $\phi-\phi_M=0.005$ we should expect a much higher nucleation rate, and crystallisation on the simulation timescale. 
To verify this point, we carried out some simulations with a monodisperse system. 
These are shown in Figs. \ref{figXtalTime}(a and b), and indeed crystallise in the regime of interest. We thus conclude that, in the regime we access, the discrepancy between experiment and simulation is likely due to the larger system size in the case of the experiments.

\subsection{Particle-level crystallisation mechanism}

Our single-particle analysis gives us the ability to shed some light
on the mechanism of crystallisation. We have noted that the metastable
fluid is dominated by the fivefold symmetric 10B cluster. In Fig.
\ref{figTimeSingle}, we show the history of a single particle, throughout
a simulation for $\phi=0.55$ and $\phi=0.58$ . We see that each particle fluctuates
and is identified in a number of different structures, including local
crystalline environments. Three regimes emerge: the metastable fluid, dominated by the 10B cluster, the final crystal, and a transition regime between the two. Note that in the transition regime, although the particle is often found in amorphous structures, these are rarely 10B. This suggests that 10B-crystal transitions may be somehow suppressed. This is consistent with 
long-standing ideas that five-fold symmetry can suppress crystallisation ~\cite{frank1952} and very recent experimental ~\cite{leocmach2012} work which suggests frustration between five-fold symmetry and local crystalline order. We would thus expect that 10B clusters are rather stable (see below). During growth, however, it is possible that a crystalline surface may disrupt the five-fold symmetry in the fluid, leading to more rapid transformation between 10B and crystalline structures. Our finding of a transition regime further supports
findings that crystallisation preferentially takes place in a region of high crystal-like bond orientational order ~\cite{kawasaki2010pnas,kawasaki2010jpcm,russo2011}. 
%and structures with five-fold characters (10B and icosahedra) are frustrated with 
%the crystal-like ordering \cite{leocmach2012}. However, once crystal nuclei are formed, the %transformation from 10B clusters to fcc clusters takes place at 
%the growth front of crystals. 

We close by considering the stability of the fivefold symmetric 10B cluster. This seems to dominate the metastable fluid at densities where crystallisation occurs. In Fig. \ref{figTransitions} we show transition probabilities from the 10B cluster to various geometries.
We see there is a tendency to remain in the 10B cluster. That is, the 10B cluster shows a higher degree of stability than other clusters. 
In other words, for the nearly hard sphere fluid, the 
10B is a locally preferred structure, similar to icosahedra and related polyhedra in glass-forming systems ~\cite{steinhardt1983,jonsson1988,tomida1995,dzugutov2002,coslovich2007,malins2012wahn}. 
This is consistent with previous reports that five-fold symmetry is favoured in `hard' spheres 
~\cite{karayiannis2011,karayiannis2012,leocmach2012}. 
The fact that this ordering tendency to structures of five-fold character such as 10B appears to be enhanced 
at higher $\phi$ is intriguing, as at higher volume fractions still, `spinodal' crystallisation takes place in a small fraction of $\tau_\alpha$ as found previously ~\cite{zaccarelli2009xtal,sanz2011}. In such unstable fluids, however, we cannot measure the lifetime of 10B clusters: the stability of 10B is exceeded 
by the thermodynamic driving force of crystal nucleation.

\section{Conclusion}

We have carefully matched simulation to experiment for a nearly hard
sphere system. For the regimes in which we can access crystallisation,
the kinetics are similar in both simulation and experiment with the
exception that, at lower volume fraction, experiments crystallise faster than simulations.
We believe these are associated with the onset of 
low nucleation rates as the supersaturation is decreased. Under these conditions, the larger experimental system size means nucelation events occur on accessible timescales, enabling
crystallisation to be observed in experiments but not in simulations. While accurate simulation data has been obtained for monodisperse systems across a wide range of $\phi$, the effect of polydispersity  characteristic of colloidal experiments has only been considered for nucleation rates too low for confocal microscopy experiments to access. In order to be confident that no discrepancy exists, predictions for polydisperse systems in the regime accessible to experiments such as ours would be helpful. 

%While the magnitude of the discrepancy in our experiments is much smaller
%than previous comparisons for data which accessed much lower nucleation rates
%\cite{auer2001}, the trends are the same. We believe that our brute force simulations
%should give reasonable estimates of the crystallisation time at these supersaturations %\cite{filion2011}. 
%Moreover the discrepancy appears to lie outside our margin of error ($\lesssim1$\% volume %fraction)
%Thus we conclude that the discrepancy between simulation and experiment in the nucleation %rate of colloidal hard spheres continues.

Our topological cluster classification reveals the mechanism of crystallisation.
In particular, around the freezing transition, nearly hard sphere
fluids become dominated by a five-fold symmetric ten-membered cluster
which we term 10B. By considering particle histories, we find that
transitions between this 10B cluster and crystalline environments
are suppressed. Instead, after some time in a metastable fluid state,
with occasional excursions to a crystalline environment, which usually
occur through an intermediate structure, the particle finds itself
in a transition state, presumeably due to the proximity of a crystalline region
which stabilises local crystalline environments. In the transition
state, the particle spends large amounts of time in a crystalline
environment, and almost no time in a 10B cluster, instead it is found in other amorphous clusters.
Eventually, the particle
spends all its time in a local crystalline environment and is said
to be crystalline. The transition regime we observe may be related to the `cloud' 
identified in the case of softened particles ~\cite{lechner2011}.

Finally, we emphasise that, since absolutely hard spheres are not found
in nature ~\cite{royall2012myth}, it is essential to take account of the inherent softness
in any experimental system. However, comparison with true hard spheres
suggests that the main effect of the softness we have considered is
to shift the state point such that we must consider effective volume
fractions. This is consistent with previous observations that mapping the effective packing fraction to
hard spheres results in a practically identical fluid structure \cite{taffs2010jcp}.

\section*{Acknowledgements}

It is a pleasure to thank Kurt Binder, Bob Evans, Daan Frenkel, Rob Jack, Mathieu Leocmach, Thomas Palberg, John Russo, 
Richard Sear, and Frank Schrieber for stimulating discussions and Monica Moreno for preliminary simulations and experiments.
CPR gratefully acknowledges the Royal Society for financial support and EPSRC grant code EP/H022333/1 
%JT is supported by EPSRC EP/E025377/1. 
H.T. acknowledges a grant-in-aid from the 
Ministry of Education, Culture, Sports, Science and Technology, Japan and 
Aihara Project, the FIRST program from JSPS, initiated by CSTP. 

%\bibliographystyle{apsrev}
%\bibliography{xtal}

\end{document}